\documentclass[twocolumn, prd, reprint, nofootinbib,showpacs,preprintnumbers,amsmath,amssymb]{revtex4}
\usepackage{varioref,exscale,latexsym,amsmath,amssymb}
\usepackage{graphicx}

\usepackage{slashed}
\usepackage{dcolumn}
\usepackage{bm}
\usepackage{hyperref}
\usepackage{color}
\usepackage{xcolor}

\newcommand{\beq}{\begin{equation}}
\newcommand{\eeq}{\end{equation}}
\newcommand{\bea}{\begin{eqnarray}}
\newcommand{\eea}{\end{eqnarray}}

\newcommand{\Tr}{\mathop{\rm Tr}}

\def\lsi{\raise0.3ex\hbox{$<$\kern-0.75em\raise-1.1ex\hbox{$\sim$}}}
\def\gsi{\raise0.3ex\hbox{$>$\kern-0.75em\raise-1.1ex\hbox{$\sim$}}}

\def\beq{\begin{equation}}

\def\eeq{\end{equation}}

\def\beqa{\begin{eqnarray}}

\def\eeqa{\end{eqnarray}}

\begin{document}
\preprint{ACFI-T20-10}

\title{{\bf The cosmological constant and the use of cutoffs}}

\medskip\

\medskip\

\author{John F. Donoghue}
\email{donoghue@physics.umass.edu}
\affiliation{
Department of Physics,
University of Massachusetts,
Amherst, MA  01003, USA\\}

\begin{abstract}
Of the contributions to the cosmological constant, zero-point energy contributions scale as $\delta^4(0)\sim \Lambda^4$ where $\Lambda$ is an ultraviolet cutoff used to regulate the calculations. I show that such contributions vanish when calculated in perturbation theory. This demonstration uses a little-known modification to perturbation theory found  by Honerkamp and Meetz and by Gerstein, Jackiw, Lee and Weinberg which comes into play when using cutoffs and interactions with multiple derivatives, as found in chiral theories and gravity. In a path integral treatment, the new interaction arises from the path integral measure and cancels the $\delta^4(0)$ contributions. This reduces the sensitivity of the cosmological constant to the high energy cutoff, although it does not resolve the cosmological constant problem. The feature removes one of the common motivations for supersymmetry. It also calls into question some of the results of the Asymptotic Safety program. Covariance and quadratic cutoff dependence are also briefly discussed.
\end{abstract}
\maketitle

\section{Cutoffs and zero-point energy}
In regularizing quantum field theories, dimensional regularization is the most common and useful choice, partially because it preserves all the symmetries of the theory. However, cutoffs also plays a role in our thinking about physics. Part of this is the legacy of the history of cutoff regularization. But there is also some genuine physics involved. We think of effective field theories as being valid up to some energy scale, and a cutoff can parameterize the limit of validity of the effective field theory. In addition, running couplings depend on the energy scale and cutoffs are sometimes used in their description. But if we are to use cutoffs, our thinking should be aligned with the underlying calculations. In this paper, I describe how direct calculations of the cosmological constant using a cutoff differ from our common description, and show the need for a new interaction term when using cutoffs with gravity.

In discussing the cosmological constant problem, we note that $\Lambda_{cc}$ corresponds to the vacuum energy density, for which there are many contributions. One which is normally mentioned is the zero-point energy. When calculated for a scalar field, using canonical quantization one writes
\beq
E_0 = \int \frac{ d^3 p}{(2\pi)^3} \frac12 \omega_p \sim \frac1{16\pi^2} \Lambda^4
\eeq
where in the second form I have cutoff the divergent momentum integral at a scale $\Lambda$. (Unfortunately, the standard convention is to call both the vacuum energy and the cutoff by the symbol $\Lambda$. I will always put the $cc$ subscript on the cosmological constant, i.e. $\Lambda_{cc}$). Since the measured value of the cosmological constant is $\Lambda_{cc}\sim (10^{-3}~{\rm eV})^4$ and we might trust the zero-point energy calculation up to the Planck mass, this leads to the common complaint about this being the ``worst prediction ever - failing by 120 orders of magnitude''. One of the motivations for supersymmetry is to cancel these effects by having equal numbers of boson and fermion degrees of freedom.

This calculation is inadequate, as it is not covariant. Indeed if we calculate all the components of the energy momentum tensor using canonical quantization, we find the $\Lambda^4$ contribution to the vacuum values is
\beq
T_{\mu\nu}|_0 = diag(1,\frac13,\frac13,\frac13)\times \frac1{16\pi^2} \Lambda^4
\eeq
such that this divergent part of the vacuum value is traceless, $\eta^{\mu\nu}T_{\mu\nu}|_0 =0$. Since the contribution to the cosmological constant can equally be identified with the trace of the energy momentum tensor
\beq
T^\mu_\mu = 4\Lambda_{cc}  \ \ ,
\eeq
we could equally well conclude that this contribution to the cosmological constant is zero. The canonical quantization calculation of the zero-point energies and momenta is not compatible with Lorentz invariance of the vacuum. The point is that covariance requires an effect proportional to $\eta_{\mu\nu}$.

\begin{figure}[htb]
\begin{center}
\includegraphics[height=30mm,width=80mm]{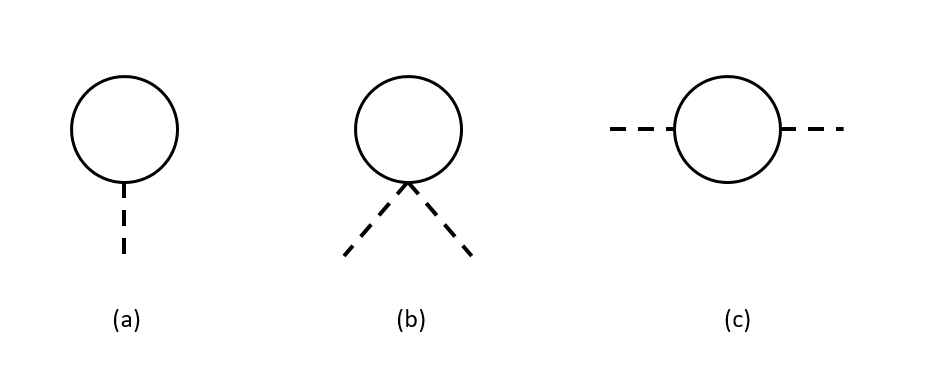}
\caption{The tadpole diagrams (a) and (b) and the bubble diagram (c). The solid line is the scalar field and the dashed line is the metric field. }
\label{tadpolediagram}
\end{center}
\end{figure}

The covariance problem can be resolved by using quantum field theory to calculate the contribution to the cosmological constant. The cosmological constant appear in the gravitational action as
\beqa
S_{grav}&=& \int d^4x ~\sqrt{-g}\left[-\Lambda_{cc} +\frac{2}{\kappa^2} R+...\right] \nonumber \\
&=& \int d^4x ~\left[-\Lambda_{cc}\left(1+\frac12 \eta^{\mu\nu}h_{\mu\nu}\right)+...\right]
\eeqa
where the expansion of $g_{\mu\nu}=\eta_{\mu\nu} + h_{\mu\nu}$ has been used, and $\kappa^2 =32\pi G$. The coupling of $h_{\mu\nu}$ to matter has the form
\beq
{\cal L}_{int} =- \frac12 h_{\mu\nu} T^{\mu\nu}   \ \ .
\eeq
To first order in $h_{\mu\nu}$ the contribution to the cosmological constant then can be identified via the tadpole loop diagram which is illustrated in Fig. \ref{tadpolediagram}a. Using a minimally coupled scalar particle in the loop we find the effect
\beqa
\Delta {\cal L} &=& -i \frac12 h_{\mu\nu} \times \int \frac{d^4p}{(2\pi)^4}\frac{2p^\mu p^\nu -\eta^{\mu\nu}(p^2-m^2)}{p^2-m^2+i\epsilon}\nonumber \\
&=&  -i\frac12 h_{\mu\nu} \eta^{\mu\nu}  \int \frac{d^4p}{(2\pi)^4}\frac{\frac12 p^2 -(p^2-m^2)}{p^2-m^2+i\epsilon} \nonumber  \\
&=& i\frac12 h_{\mu\nu} \eta^{\mu\nu} \frac12 \delta^4(0) +{\cal O}(m^2)\nonumber \\
&\sim& -\frac12 h_{\mu\nu} \times \eta^{\mu\nu}  \frac1{64\pi^2} \Lambda^4
\eeqa
yielding the covariant definition of this contribution
\beq
\Lambda_{cc} \sim \frac1{64\pi^2} \Lambda^4  \ \ .
\eeq

However, in this paper I will show that even the above QFT calculation is wrong - that in fact there are no perturbative $\delta^4(0)\sim \Lambda^4$ zero-point contributions to the cosmological constant. The demonstration is an outgrowth of a little-known feature about the use of perturbation theory when using cutoffs in situations where the interactions are proportional to the derivatives of fields. In the context of chiral theories, Honerkamp and Meetz \cite{Honerkamp:1996va}  and Gerstein, Jackiw, Lee and Weinberg \cite{Gerstein:1971fm} (GJLW) showed that theories with derivative-based interactions have a new ingredient in the Feynman rules that is relevant when using cutoffs\footnote{The new effect vanishes in dimensional regularization, which is one reason that it is little known today.}. In a covariant path integral treatment, that interaction is part of the path integral measure. This exactly cancels the quartic $\Lambda$ dependence in loops. A similar effect appears in gravitational interactions of matter fields and in pure gravity. In pure gravity, this was shown by Fradkin and Vilkovisky \cite{Fradkin:1974df}\footnote{I thank A. Tseytlin for bringing this reference to my attention after the first version of the preprint was posted. I comment on the slight difference between their result and mine in Section \ref{gravity}.}. The new interaction is proportional to $i\delta^4(0)\log \det( -g_{\mu\nu})$, where $\delta^4(0)$ is $\delta^4(x)$ evaluated at $x=0$, and is relevant in any regularization scheme where the regularized value of $\delta^4(0)$ is non-zero.

\section{Scalar fields- simplified metric}

Here let us consider a massless scalar field, with the action
\beq
S= \int d^4x \sqrt{-g} ~\frac12 g^{\mu\nu} \partial_\mu \phi \partial_\nu \phi  \ \ .
\eeq
The mass term will be discussed separately below.

\subsection{Simplified metric}

In this section, I will illustrate the physics involved using a simplified metric.
\beq
g_{\mu\nu}(x) = \left(1+h(x)\right) \eta_{\mu\nu}  \ \ ,
\eeq
This choice is motivated by the fact that one can directly use the results of GJLW \cite{Gerstein:1971fm} without any need for new derivations, and that the results emerge simply. While we can keep in mind that this follows from a generally covariant action, the result can be treated in perturbation theory using ordinary Minkowski space field theory. It is just an ordinary field theory with a derivative coupling between the scalar field and the background field $h(x)$. Moreover this metric is general enough that we can isolate the cosmological constant contribution cleanly. So overall we can see the essential physics in this example, without any elaboration of the formalism. After we see the underlying physics in this pedagogic example, we can readily understand the more general cases. By general covariance, the results which are shown using this metric will also yield the same result with a more general metric.

With this choice of metric, we have the following relations
\beq
\sqrt{-g} = \left(1+h(x)\right)^2
\eeq
and
\beq\label{curvature}
\sqrt{-g}R =  \frac32 \frac{\partial_\mu h \partial^\mu h}{1+h} - 3\Box h \ \ .
\eeq
The important feature here is that we can recognize the effects of these two terms through their derivative structure. Renormalization of the vacuum energy involves the metric field $h$ with no derivatives. Renormalization of the Newton constant involves two derivatives of $h$. At the order of curvature squared, there are in general two invariants in the 4D action and we can recover only one invariant, $R^2$, because this metric satisfies $R_{\mu\nu}R^{\mu\nu}- \frac13 R^2 =0$. However, since our primary focus is on the cosmological constant, this metric is general enough for our purposes.

The massless scalar field action involves
\beqa\label{masslessaction}
\sqrt{-g}{\cal L} &=& \frac12 \sqrt{-g} g^{\mu\nu} \partial_\mu \phi \partial_\nu \phi \nonumber \\
&=& \frac12\left(1+h(x)\right) \partial^\lambda \phi \partial_\lambda \phi  \ \ .
\eeqa
The coupling of the field $h$ to the scalar involves two derivatives, and this is the feature which generates the non-standard features in the Feynman rules.

The action is has no dimensional parameters. Using a cutoff adds a mass dimension to the theory which was not there originally. If we were investigating physical observables in this theory, it would be sufficient to use dimensional regularization, which seems better suited for a massless theory. However, lets see what happens with the use of a cutoff.

\subsection{Canonical quantization}

The key new ingredient here is the fact that when the interactions contain derivatives, the canonical momentum is changed,
\beq
\Pi(x) = \frac{\partial {\cal L}}{\partial {\dot{\phi}}} = \left(1+h(x)\right){\dot{\phi}}
\eeq
When forming the interaction Hamiltonian and going to the interaction picture, this induces a new interaction.
As for the new interaction, here is a brief review of the GJLW results using their notation, after which we will revert to the special case for the gravitational interaction. The authors start with a Lagrangian for multiple fields $\pi^a (x)$
\beq
{\cal L} = \frac12 G^{ab}(\pi) \partial_\mu \pi^a \partial^\mu \pi^b
\eeq
with
\beq
G^{ab}(\pi) = \delta^{ab} +\bar{G}^{ab}(\pi)   \ \ .
\eeq
The canonical momentum is
\beq
\Pi^a(x) =\left( \delta^{ab} +\bar{G}^{ab}(\pi)\right)\partial_0 \pi^b(x)   \ \ .
\eeq
Forming the Hamiltonian yields
\beqa
{\cal H} &=& {\cal H}_0+{\cal H}_I  \nonumber \\
{\cal H}_0 &=&\frac12 \Pi^a\Pi^a + \partial_i \pi^a \partial_i \pi^a   \nonumber \\
{\cal H}_I &=&- {\cal L}_I - \frac12 \partial_0 \pi^a \bar{G}^2_{ab} \partial_0\pi^b
\eeqa
with the standard interaction Lagrangian
\beq
{\cal L}_I = \frac12 \bar{G}^{ab}(\pi) \partial_\mu \pi^a \partial^\mu \pi^b  \ \ .
\eeq
Now in going to the interaction picture we identify
\beqa
\pi^a &\to \phi^a  \nonumber \\
\Pi^a &\to \partial_0 \phi^a
\eeqa
where $\phi^a$ is the interaction picture field. The resulting perturbative Hamiltionian is
\beq
{\cal H}_I=  - \frac12 \bar{G}^{ab} \partial_\mu \phi^a \partial^\mu \phi^b + \frac12 \partial_0 \phi^a \left[\frac{\bar{G}^2}{1+ \bar{G}} \right]_{ab}\partial_0\phi^b \ \ .
\eeq
The first term yields the usual perturbative expansion. The second term is new. If we specialize the the special metric that we are studying, the equivalent form is
\beq\label{nonstandardinteraction}
{\cal H}_I = - \frac12 h \partial_\mu \phi \partial^\mu \phi + \frac12 \partial_0 \phi \left[\frac{h^2}{1+ h} \right]\partial_0\phi \ \ .
\eeq
The new term starts at second order in $h$.

In addition, the propagator functions pick up a modification when multiple derivatives are included. If we define
\beqa
\Delta(q) &=& \int d^4 x e^{iqx}\langle 0|T\phi(x)\phi(0)|0\rangle  \nonumber \\
\Delta_\mu (q) &=& \int d^4 x e^{iqx}\langle 0|T\partial_\mu \phi(x)\phi(0)|0\rangle  \nonumber \\
\Delta_{\mu\nu}(q) &=& \int d^4 x e^{iqx}\langle 0|T\partial_\mu\phi(x)\partial_\nu \phi(0)|0\rangle
\eeqa
then the first two have the usual form, but the third has a modification due to the time ordering
\beqa\label{propagators}
\Delta(q) &=& \frac{i}{q^2+i\epsilon} \nonumber \\
\Delta_\mu (q) &=& \frac{iq_\mu}{q^2+i\epsilon} \nonumber \\
\Delta_{\mu\nu}(q) &=& \frac{iq_\mu q_\nu}{q^2+i\epsilon} -i\eta_{\mu 0}\eta_{\nu 0}
\eeqa
The non-standard term in the propagator follows from the commutation rules for the interaction picture field $\phi$.
While various steps are not Lorentz covariant, the final results will be.

Let us calculate the loop contribution to the cosmological constant. The tadpole diagram with one external factor of $h$, Fig. 1a,  is now
\beq\label{tadpole}
-i{\cal M} = i \int \frac{d^4p}{(2\pi)^4}  ~ \left[\frac{p^2}{p^2+i\epsilon} -1 \right] \ \ .
\eeq
The first term is the usual Feynman rule. The second comes from the extra piece in the propagator of Eq. \ref{propagators}. One sees that they cancel.

Let us confirm this result by looking at diagrams with two factors of $h$. The treatment here is more subtle because it involves the new non-standard interaction of Eq. \ref{nonstandardinteraction}. The bubble diagram, Fig. \ref{tadpolediagram}(c), contributes
\beq
\Sigma_B (q) =\frac12 \int \frac{d^4 p}{(2\pi)^4} \left[\frac{\left(p\cdot(p-q)\right)^2}{p^2 (p-q)^2}  - 2 \frac{p_0p_0}{p^2}+1\right]
\eeq
with the first term being the standard interaction and the others coming from the propagator modification. There is also a tadpole diagram with two factors of $h$, Fig. \ref{tadpolediagram}(b), coming from the non-standard interaction which contributes
\beq
\Sigma_T = \int \frac{d^4 p}{(2\pi)^4} \left[ \frac{p_0p_0}{p^2}-1\right]   \ \ .
\eeq
These sum to
\beq
\Sigma (q) =\frac12 \int \frac{d^4 p}{(2\pi)^4} \left[\frac{\left(p\cdot (p-q)\right)^2}{p^2 (p-q)^2}  - 1\right]  \ \ .
\eeq
As expected the non-covariant terms cancel. It is easy to verify that the quartic cutoff dependence in this expression vanishes. In addition, we have
\beq
\Sigma (q=0)=0
\eeq
which implies that the interaction without derivatives vanishes. This is consistent with the vanishing of the single $h$ tadpole diagram which we found above in Eq. \ref{tadpole}.

\subsection{Field redefinition}\label{fieldredefinition}

This result can be understood without even calculating it explicitly. We can see from the action, Eq. \ref{masslessaction}, that if the metric field $h$ was a constant, the pre-factor would be absorbed into the normalization of the field $\phi$. In the wavefunction renormalization process one absorbs an overall factor into the normalization of the field
\beq
\frac12 Z \partial_\mu\phi\partial^\mu \phi\to \frac12  \partial_\mu\phi^r \partial^\mu \phi^r  \ \ .
\eeq
Even if we did not do this explicitly, the quantization process would take care of the normalization when one defines normalized states. To the extent that $1+h$ is a constant, that should happen in a consistent quantization. To the extent that $1+h$ is {\em not} a constant, the theory should depend on the derivatives of $h(x)$

One can see this explicitly through a field redefinition - specifically a local wavefunction renormalization. For a general $h(x)$ the interaction pre-factor can be removed by a field redefinition
\beq
\phi = \frac1{\sqrt{1+h}}\chi  \ \ .
\eeq
This transforms the original Lagrangian into
\beq\label{transformedaction}
{\cal L}= \frac12\left[\partial_\mu \chi \partial^\mu \chi - \chi\partial^\mu \chi \frac{\partial_\mu h}{1+h}+ \frac14 \chi^2 \frac{\partial_\mu h\partial^\mu h}{(1+h)^2}\right]  \ \ .
\eeq
In this form, we can see that all of the interactions of the metric field $h$ involve its derivatives $\partial_\mu h$. Since we are only studying loops of the scalar particle, there is no chance of removing this derivative factor. Thus loops will not generate the vacuum energy term which involves only powers of $h$ without derivatives.

This form of the action also involves a derivative interaction , which in this case is linear in the derivative of $\chi$. This is similar to scalar QED which also has a linear derivative interaction. Investigations of canonical quantization of scalar QED show that the non-standard vertex which follows from the derivative interaction gets canceled by the non-standard term in the propagator, Eq. \ref{propagators} (e.g. see \cite{Weinberg:1995mt}). We can proceed without concern in this case. Equivalently we could use integration by parts to remove the $\partial^\mu\chi$ interaction, which yields
\beq\label{transformedaction2}
{\cal L}= \frac12\left[\partial_\mu \chi \partial^\mu \chi - \frac16 \chi^2 (-g)^{1/4} R\right]  \ \ .
\eeq
where we have identified the scalar curvature for this particular metric, Eq. \ref{curvature}. In this form it is clear that the cosmological constant term will not be generated by loops of the scalar field.

\subsection{Covariant quantization}

Canonical quantization is awkward for this problem, as it leads to non-covariant interactions and propagators. However, there exists a covariant version also, in which one uses only the usual covariant propagators. It however involves a new interaction term which accounts for the effects of both the non-standard propagators and the non-standard interaction discussed above. In the notation of GJLW, this has the form
\beq
\Delta {\cal L}= -\frac12 i\delta^4(0) ~\Tr \log \left[1 + \bar{G}(\phi)\right]  \ \ .
\eeq
Here $\delta^4(0) $ is to be identified with
\beq
\delta^4(0) = \int \frac{d^4p}{(2\pi)^4} = i \frac1{32\pi^2} \Lambda^4
\eeq
when using cutoff regularization. This factor vanishes in dimensional regularization as it is a scaleless integral. For our simplified metric, one has
\beq
\Delta {\cal L}= -\frac12 i\delta^4(0) ~\log \left[1 + h(x)\right]  \ \ .
\eeq

The GJLW derivation is direct in accounting for the non-standard features starting from the Hamiltonian formalism. However it is somewhat convoluted. For the present case, it is easier to mimic the technique of Honerkamp and Meetz \cite{Honerkamp:1996va} (see also DeWitt \cite{DeWitt:1962ud}, Boulware \cite{Boulware:1970zc} and Salam and Strathdee  \cite{Salam:1971sp}) . This follows from the observation of subsection \ref{fieldredefinition} that in the representation of Eqs. \ref{transformedaction},  \ref{transformedaction2} there is no need for any alternate rules. Using the path integral in Lagrangian form, one can then transform back from the field variables $\chi$ to $\phi$. In the path integral, this comes with a Jacobian factor
\beq
\int \left[d \chi\right] = \int \left[d \phi\right] J \ \ .
\eeq
Here we start with the transformation
\beq
D(x,x') = \frac{\delta \chi (x)}{\delta \phi(x')} = \delta^4(x-x') \frac{\partial \chi}{\partial \phi}(x)
\eeq
Because there are no derivatives involved, the Jacobean factor is local in the
coordinate representation
\beq\label{jacobian}
J = {\rm Det}\left[ D(x,x)\right]
\eeq
where Det implies the functional determinant. Exponentiation of this factor is then accomplished via
\beq
J= {\rm Det} D = e^{\Tr \log D}  \ \ .
\eeq
Evaluating this we find
\beq
\log D(x,x') = \delta^4(x-x')\log \sqrt{1+h(x)}  \ \ ,
\eeq
such that
\beq
J= {\rm Det} D = e^{\delta^4(0)\int d^4x \log \left[ \left(1+h\right)^{1/2}\right]} \ \ .
\eeq
The $\delta^4(0)$ factor arises from the locality in $x$ of Eq. \ref{jacobian}. This yields the new interaction in the Lagrangian.

Use of this interaction then accounts for the removal of the $\Lambda^4$ factors that were previously accomplished by the non-standard Hamiltonian interactions. Once can see this most simply in the linear term in $h$, which amounts to
\beq
\frac12 i \delta^4(0)  = - \frac{1}{64\pi^2}\Lambda^4
\eeq
cancelling the quartic contribution to the cosmological constant.

\subsection{Quadratic cutoff dependence and masses}

This demonstration should not be taken to mean that cutoffs cannot lead to the renormalization of the cosmological constant. When the particle's mass is not zero, there will be a quadratic cutoff dependence in the renormalization
\beq
\delta \Lambda_{cc} \sim m^2 \Lambda^2  \ \ .
\eeq
This can be seen simply in our calculation of the tadpole diagram. The Lagrangian with a mass term is
\beqa\label{withmass}
\sqrt{-g}{\cal L} &=& \frac12 \sqrt{-g}\left[ g^{\mu\nu} \partial_\mu \phi \partial_\nu \phi - m^2 \phi^2 \right]
 \\
&=&\frac12 \left[\left(1+h(x)\right) \partial^\lambda \phi \partial_\lambda \phi - m^2 \left(1+h(x)\right)^2\phi^2 \right]\nonumber
\eeqa
The tadpole diagram now becomes
\beqa\label{massivetadpole}
-i{\cal M} &=& i \int \frac{d^4p}{(2\pi)^4}  ~ \left[\frac{p^2-2m^2}{p^2 -m^2+i\epsilon} -1 \right] \nonumber \\
&=& i \int \frac{d^4p}{(2\pi)^4}  ~ \left[\frac{-m^2}{p^2 -m^2+i\epsilon} \right]\ \ .
\eeqa
The last integral is quadratically divergent.

The renormalization of cosmological constant can also be seen using the field transformed version of the action. Including a mass term has the effect that the rescaling of fields from $\phi\to\chi$ now leaves behind a non-derivative interaction. After the rescaling of the field, we find
\beqa\label{massiveaction}
\sqrt{-g}{\cal L} &=& \frac12\partial_\mu \chi \partial^\mu \chi -\frac12  \chi\partial^\mu \chi \frac{\partial_\mu h}{1+h} \nonumber \\
&+& \frac18 \chi^2 \frac{\partial_\mu h\partial^\mu h}{(1+h)^2}- \frac12 m^2 (1+h)\phi^2 \ \ .
\eeqa
In the loop diagrams there will now be a residual non-derivative interaction proportional to $m^2\Lambda^2$.

\section{Scalars - general metric}

The analysis of the previous section contains the ingredients for the treatment of a general metric. For any metric, we can always divide the metric into an overall conformal factor and a unimodular metric
\beq\label{metric}
g_{\mu\nu}(x) = \Omega^2(x)\bar{g}_{\mu\nu}
\eeq
with
\beq
\det \bar{g}_{\mu\nu}= -1  \ \ .
\eeq
In this division, we have
\beq
- \det g_{\mu\nu} = \Omega^8 \ \ .
\eeq
The scalar field Lagrangian is now
\beqa
S&=& \int d^4x \Omega^2 ~\frac12 \bar{g}^{\mu\nu} \partial_\mu \phi \partial_\nu \phi \nonumber \\
&=& -\int d^4x \Omega^2 ~\frac12 \phi D^2\phi \ \ ,
\eeqa
with
\beq
D^2 = \partial_\mu \bar{g}^{\mu\nu} \partial_\nu + \frac{2}{\Omega}\left(\partial_\mu \Omega \right) \bar{g}^{\mu\nu} \partial_\nu \ \ .
\eeq

The goal of our study, the cosmological constant term, is given uniquely by the conformal factor
\beq
\int d^4 x \sqrt{-g} \Lambda_{cc} = \int d^4 x ~\Omega^4 \Lambda_{cc}  \ \ .
\eeq

The correspondence with the specialized metric is
\beq
1 +h(x) \to \Omega^2  \ \ .
\eeq
The new interaction becomes
\beqa
\Delta {\cal L}&=& -\frac12 i\delta^4(0) ~\log \left[1 + h(x)\right] \nonumber \\
&\to& -i\delta^4(0) ~\log \left[\Omega)\right] \nonumber \\
&=&-i\frac18 \delta^4(0) ~\log \left[-\det g_{\mu\nu}\right]  \ \ .
\eeqa
Here $\det$ refers only to the determinant over the Lorentz indices, as in $g= \det g_{\mu\nu}$.

We can verify that this factor removes the contribution to the cosmological constant through a path integral treatment. The path integral with the Jacobean included is
\beq
\int [d\phi] {\rm Det}( -g)^\frac18 e^{i \int d4 x \sqrt{-g} {\cal L}} = \int [d\phi] ({\rm Det \Omega}) e^{-i \int d4 x ~\Omega^2 \phi D^2 \phi} \ \ .
\eeq
Within $D^2$ there are only derivative interactions of $\Omega$ so that these terms cannot lead to a change in the cosmological constant.
Doing the path integral we obtain
\beq\label{scalardeterminant}
({\rm Det \Omega}) \frac1{ ({\rm Det }\Omega^2 D^2)^{\frac12}} =\frac1{({\rm Det } D^2)^{\frac12}} \ \ .
\eeq
Here we see how the path integral accomplishes the local wavefunction renormalization - the overall factors of $\Omega$ cancel. There is no modification of the cosmological constant term.

\section{Fermions}

There is a related discussion for fermions. First we can examine the situation with the simplified metric.
The massless action involves
\beq
\sqrt{-g} {\cal L} = (1+h)^{\frac32}\bar{\psi} \left[i \gamma^\mu\left({\partial}_\mu-\frac{1}{8(1+h)}\sigma_{\mu\nu}(\partial^\nu h) \right) \right] \psi
\eeq
where the second term is the spin connection term in this metric..
The spin connection term in the Lagrangian involves the derivative of the metric, so that we will not display this piece when discussing the cosmological constant.
The propagator rules which we need are
\beqa
iS(q) &=& \int d^4 x e^{iqx}\langle 0|T|\psi(x)\bar{\psi(0)}|0\rangle  \nonumber \\
iS_\mu (q) &=& \int d^4 x e^{iqx}\langle 0|T\partial_\mu {\psi(x)}\bar{\psi(0)}|0\rangle
\eeqa
with the representation
\beqa\label{fermionpropagators}
iS(q) &=& \frac{i}{\slashed{q}+i\epsilon} \nonumber \\
iS_\mu (q) &=& \frac{q_\mu}{\slashed{q}+i\epsilon} - \gamma_0\delta_{\mu 0}
\eeqa
The unconventional extra term follows from the commutation rules for fermions.

From the Lagrangian and the propagators we can calculate the loop correction to the cosmological constant by looking at the linear term in $h(x)$ via the tadpole diagram of Fig. 1a. One obtains
\beq
\delta {\cal L} = \frac32 i h \int \frac{d^4q}{(2\pi)^4} \left[\frac{\slashed{q}}{\slashed{q}+i\epsilon}-1\right] \ \ .
\eeq
As before this gives a null result for the $\Lambda^4$ contribution to the cosmological constant. (Adding a mass will lead to quadratic cutoff dependence.)

When we go to a generic metric, we can again use the decomposition of Eq. \ref{metric}. The Lagraingian is now
\beq
\sqrt{-g}{\cal L} = \Omega^3 \bar{\psi} \left[i \bar{e}^\mu_a\gamma^a {\partial}_\mu+\frac14\sigma_{bc}A^{bc}_\mu \right] \psi
\eeq
where $\bar{e}^\mu_a$ is the vierbein for the unimodular metric and
\beq
A^{bc}_\mu = \bar{A}^{bc}_\mu +\frac{1}{\Omega} \left(e^b_\mu e^{a\lambda}- e^a_\mu e^{b\lambda}\right)\partial_\lambda \Omega
\eeq
is the spin-connection, with $\bar{A}^{bc}_\mu$ being the spin connection formed using the unimodular metric with no factors of $\Omega$. Again the spin connection contains derivatives of the metric and can be neglected for our purposes.
The conformal factor without derivatives can be removed from the action by the local wave function renormalization
\beq
\eta (x) = \Omega^\frac34 \psi \ \ ,
\eeq
which allows us to use conventional Feynman rules and agree with the non-renormalization of the cosmological constant. By our previous work this leads to the appropriate measure
\beq
d\psi d\bar\psi \det \left[\Omega\right]^\frac32 = d\psi d\bar\psi \left[-\det g_{\mu\nu}\right]^\frac38  \  \ .
\eeq
Exponentiating this leads to the novel interaction term
\beq
\Delta {\cal L} = -i \frac38 \delta(0) \log (- g )  \ \ .
\eeq

\section{Photons}

With photons the vanishing of the contribution to the cosmological constant is most easily seen by going directly to the general metric. In this case,
\beq
\sqrt{-g}{\cal L} = \Omega^4 \Omega^{-4} \bar{g}^{\mu\alpha} \bar{g}^{\nu\beta} F_{\mu\nu}F_{\alpha\beta}
\eeq
so that the conformal factor totally disappears from the action. In the gauge fixing term
\beq
\sqrt{-g} {\cal L}_{g.f.} = \sqrt{-g} \frac{1}{2\xi}\left(g^{\mu\nu}D_\mu A_\nu\right)^2  \ \ .
\eeq
the overall factor of $\Omega$ also vanishes. The connection only contains derivatives of the metric. These will generate only powers of the curvatures, which are formed using derivatives of the metric. Given that the 4D action of photons coupled to gravity does not contain any overall factors of $\Omega$, there can be no shift in the cosmological constant from photon loops in any 4D regularization scheme.

\section{Gravitons}\label{gravity}
This analysis can be extended to gravitons themselves by use of the background field method. In this case we use the metric
\beq
g_{\mu\nu} = \Omega^2 \bar{g}_{\mu\nu} + \kappa h_{\mu\nu} = \tilde{g}_{\mu\nu}+ \kappa h_{\mu\nu}
\eeq
where $h_{\mu\nu}$ is the quantum field, $\kappa^2 = 32\pi G$, and again $\bar{g}_{\mu\nu}$ is unimodular. If we work in the harmonic gauge, the Einstein Lagrangian becomes
\beqa
\sqrt{-g} \frac{2}{\kappa^2} R &=& \frac{2}{\kappa^2}\Omega^4\tilde{R} + \Omega^{-2}\left[2 \tilde{R}_{\mu\nu}h^{\mu\nu}h^\nu_\alpha- \tilde{R}_{\mu\nu}h^{\mu\alpha}h^\alpha_\alpha \right. \nonumber \\
&+&\left.\frac12 h_{\mu\nu , \alpha} h^{\mu\nu , \alpha}-\frac14 h^\mu_{\mu , \alpha} h_\nu^{\nu , \alpha}\right]
\eeqa
up to quadratic order in $h_{\mu\nu}$. Here the displayed indices are now raised and lowered with the unimodular metric while the covariant derivatives are defined using the full background metric $\tilde{g}_{\mu\nu}$.

At this stage, the quantum part of the metric $h_{\mu\nu}$ is an ordinary quantum field. The various aspects of our treatment of the matter fields can readily be repeated.  In the canonical treatment, there is the propagator modification
\beq
\Delta^{\mu\nu}_{\alpha\beta\gamma\delta}(q) = \int d^4 x e^{iqx}\langle 0|T\partial^\mu h_{\alpha\beta}(x)\partial^\nu h_{\gamma\delta}(0)|0\rangle
\eeq
which in harmonic gauge becomes
\beq\label{noncovprop}
\Delta^{\mu\nu}(q)_{\alpha\beta\gamma\delta} = P_{\alpha\beta\gamma\delta}\left[ \frac{iq^\mu q^\nu}{q^2+i\epsilon} -i\eta^{\mu 0}\eta^{\nu 0}\right]
\eeq
where
\beq
P_{\alpha\beta\gamma\delta} =\frac12\left[\eta_{\alpha\gamma} \eta_{\beta\delta}+\eta_{\alpha\delta} \eta_{\beta\gamma}- \eta_{\alpha\beta} \eta_{\gamma\delta}\right]  \ \ .
\eeq
Using the simplified metric one can see that the renormalization of the cosmological constant vanishes in the same way as with matter fields. In the covariant method the new interaction is
\beq
\Delta {\cal L} = i \frac18 \delta(0) \log (- \tilde{g} )  \ \ .
\eeq

The ghost Lagrangian involves a fermionic vector field, $\eta_\mu$. The action has no overall factors of $\Omega$ and therefor only contains derivative interactions
\beq
\sqrt{-g}{\cal L}_{gh} = \sqrt{-g}\eta^{*\mu}\left(\eta_{\mu,\lambda}^\lambda - R_{\mu\nu}\eta^\nu\right) \to \eta^{*\mu}\left(\eta_{\mu,\lambda}^\lambda - \tilde{R}_{\mu\nu}\eta^\nu\right)
\eeq
where in the last expression the indices are raised and lowered with the unimodular metric. Much like the photon case, loops of the ghosts can only generate renormalization of curvatures which contain derivatives of the metric, and not the cosmological term.

Fradkin and Vilkovisky \cite{Fradkin:1974df} have also shown, using different methods, that the $\delta^4(0) $ self-energy effects are cancelled by a measure interaction in pure gravity. Their result is slightly different, containing $g_{00}$ in addition to the determinant $g$. This can be traced back to the fact that they continue to use a non-covariant propagator similar to Eq. \ref{noncovprop}. However, it is clear that their measure interaction is designed to have the same net effect, and would be covariant if they used covariant propagators.

\section{Power-law cutoffs and general covariance}\label{covariance}

Momentum space cutoffs don't respect diffeomorphism invariance. Given the possibility of rescaling coordinates and momenta, it becomes ill-defined in which coordinate system the cutoff should be applied. Under the rescaling, the value of the cutoff will change.

There is an example which can be extracted from the work above which illustrates this. The transformed interaction used above in subsection \ref{fieldredefinition} can be put in the form of Eq. \ref{transformedaction2}.
Note the $(-g)^{1/4} $ instead of $\sqrt{-g}$.

At one loop, the tadpole diagram will yield a quadratic cutoff for the scalar curvature, of the form
\beq
\sim \Lambda^2  (-g)^{1/4} R
\eeq
from the tadpole diagram. Here the relevant integral leading to a factor of $\Lambda^2$ is
\beq
\int \frac{d^4k}{(2\pi)^4} \frac{1}{k^2} = \frac{i}{16\pi^2}\Lambda^2  \   \ .
\eeq
This interaction would not be covariant because of the $ (-g)^{1/4} $ factor\footnote{This term would vanish in dimensional regularization.}. This tells us that with this choice of field variable (itself being rescaled), one should build in a corresponding rescaling of $\Lambda$. The rescaling of $\Lambda$ which works is to replace the quadratic dependence by
\beq\label{rescaling}
\Lambda^2 \to \Lambda^2 (-g)^{1/4} \ \ .
\eeq
This can also be seen in the renormalization of the cosmological constant which occurs with a mass. We derived above the form of
Eq. \ref{massiveaction}
\beqa\label{massiveaction2}
\sqrt{-g}{\cal L} &=& \frac12 \partial_\mu \chi \partial^\mu \chi - \frac12 \chi\partial^\mu \chi \frac{\partial_\mu h}{1+h}  \nonumber \\
 &+& \frac18 \chi^2 \frac{\partial_\mu h\partial^\mu h}{(1+h)^2}- \frac12 m^2 (1+h)\phi^2  \ \ .
\eeqa
This was used to argue that a non-derivative residual effect was possible. However, if you actually do the calculation, you get a non-covariant cutoff dependence. This is readily seen, as the mass term comes with the factor $(1+h)$, so that after calculating the tadpole loop one gets an effect proportional to
\beq
\sim m^2 (1+h) \Lambda^2 \sim m^2\Lambda^2  (-g)^{1/4}
\eeq
Again a rescaling of $\Lambda$ would be needed to restore covariance.

At second order in the curvature, we get the expected covariant form
\beq
\log \Lambda^2 ~\sqrt{-g}R^2 = \log \Lambda^2~\left[ (-g)^{1/4} R\right]^2
\eeq
from the bubble diagram. This is the physical term which we would also get in dimensional regularization\footnote{The bubble diagram also leads to a $R \log \Box R$ interaction, and $\log(-g)$ from the rescaling of $\Lambda$ become part of the covariant version of $\log \Box$. }. (Recall that one cannot distinguish $R^2$ from $R_{\mu\nu}R^{\mu\nu}$ using this metric because it is conformally flat and the Weyl tensor vanishes.)

The point is not just that one can restore covariance by this simple trick - that feature is a function of these field variables. The larger issue is that cutoffs can be suspect in General Relativity because of the lack of general covariance. These calculations provide a simple example of this difficulty.

It is also correct that the new interaction required
\beq
\Delta {\cal L}= -i\frac18 i\delta^4(0) ~\log (-g) \ \ .
\eeq
is not generally covariant. However, this is designed to cancel off a related $\Lambda^4$ dependence and it disappear from the final results.

\section{Discussion}

Quantum field theory in a gravitational background provides a technique for discussing the renormalization of the terms in the effective gravitational action from loops of matter fields.  I have shown that a careful treatment of free fields coupled to gravity shows that there are no $\delta^4(0) \sim \Lambda^4$ contributions to the cosmological constant when using a cutoff regularization. To implement this in a covariant fashion requires a new interaction proportional to
\beq
i \delta^4(0) \log( -g  ) \ \ .
\eeq
This vanishes in dimensional regularization, but is important in any cutoff scheme where the regularized value of
\beq
\delta^4(0) = \int \frac{d^4k}{(2\pi)^4}
\eeq
is non-zero. The effect of the interaction is to cancel off the naive $\Lambda^4$ contribution to the cosmological constant.

Technically, this effect comes about because a constant background field, $1+h$ or $\Omega$, can be absorbed into the wavefunction renormalization of the fields, or equivalently the normalization of the states. Non-constant background fields then manifest themselves through derivative interactions. In our examples, this was implemented by a local field renormalization, with residual derivative interactions. The path integral with the new interaction naturally exhibits this feature.

This observation will not influence the calculation of physical observables, as they are expressed in terms of the physical renormalized values of the cosmological constant and the Newton constant, and are independent of regularization scheme.

However, it can influence our thinking. We often use the cutoff as a way to represent the limit of our understanding of the effective field theory. For gravity by itself, we might hope that this could be used up to near the Planck scale. A $\Lambda^4$ contribution to the vacuum energy would then be enormous. Removing this is often invoked as a motivation for supersymmetry. Having equal numbers of bosonic and fermionic degrees of freedom would lead to a cancelation of the $\Lambda^4$ contributions. The motivation for this cancelation disappears if the zero-point energies do not generate $\delta^4(0)\sim \Lambda^4$ effects.

The vanishing of the $\delta^4(0)$ contributions does not mean that the cosmological constant is not shifted by radiative corrections. There are many other contributions. Using a cutoff, there are still zero-point effects of order $m^2\Lambda^2$ which individually do not vanish. There can even be $\Lambda^4$ effects by combining two $\Lambda^2$ effects. For example\footnote{I thank D. Harlow for suggesting this particularly simple example}, in $\lambda \phi^4$ theory there can be a vacuum energy contribution at two loop order with each of the loops generating a result proportional to $\Lambda^2$. This then poses a different problem for the model builder who wishes to avoid net $\Lambda^4 \sim M_P^4$ effects, as it would require the $\lambda \phi^4$ interaction of the Standard Model to be generated dynamically at a lower scale.

Cutoffs are often used as proxies for energies in running couplings. For example, in the present practice of Asymptotic Safety\cite{Niedermaier:2006wt, Reuter:1996cp, Reuter:2019byg, Codello:2006in, Donoghue:2019clr}, one uses a cutoff to separate the high and low energy regions of the Euclidean momentum space. In contrast to common use, the cutoff is used in the infrared to remove low energy quantum effects.  By varying that cutoff, running couplings are defined. In the version of the theory where the action is truncated to the cosmological constant and the Einstein term, the one-loop running of the cosmological constant is given by\footnote{The Asymptotic Safety literature does not use the vacuum energy $\Lambda_{cc}$ as a parameter, but rather a combination $G\Lambda_{cc}$. However the separate beta functions are easy to disentangle, leading to Eq. \ref{runningvac}.  }
\beq\label{runningvac}
\Lambda \frac{\partial}{\partial \Lambda} \Lambda_{cc} = - \frac1{4\pi^2} \Lambda^4  \ \ .
\eeq
Because this is an infrared cutoff, the experimental value is obtained when the cutoff is removed ($\Lambda\to 0$) such that the solution is
\beq
 \Lambda_{cc} = \Lambda_{cc}|_{expt} - \frac1{16\pi^2} \Lambda^4  \ \
\eeq
with the UV fixed point for $\Lambda_{cc}/\Lambda^4$ being infinitely anti-deSitter space.
This running comes from the tadpole diagram of Fig. 1(a). Here I have changed the notation from describing the cutoff scale by the symbol $k$ common in the AS community to the notation used in the present paper, and to be precise this beta function reflects the cutoff function use in Ref. \cite{Codello:2006in}. There is also related $\Lambda^4$ running due to matter fields \cite{Codello:2011js}. While I have shown \cite{Donoghue:2019clr} that this is not really a running coupling in the usual sense\footnote{Briefly this is because the infrared quantum physics needs to be adding in when calculating physical observables, which would remove the $\Lambda$ dependence, and the one-loop $\Lambda$ dependence does not relate to any physical kinematic variable.}, the work of the present paper suggests that the $\Lambda^4$ dependence is not correct. There can in principle be such dependence by combining two $\Lambda^2$ effects at higher loop order, but it is clear that the original calculations done without the measure interaction need to be reconsidered. Moreover, the larger Asymptotic Safety program has yet more than two derivatives in the kinetic energy and interaction terms, so that a further modification of the Feynman rules may be needed for higher order truncations.

In general, cutoffs are problematic for general relativity because they do not respect the coordinate invariance of the theory. However when using any regularization scheme for which the scaleless integral $\delta^4(0)$ does not vanish, there needs to be a new interaction in the path integral measure which has the effect of cancelling these singular terms.

\section*{Acknowledgements}

I would like to thank Lorenzo Sorbo, Barry Holstein, Arkady Tseytlin, Daniel Harlow  and Roberto Percacci for helpful comments. This work has been partially supported by the US National Science Foundation under grant NSF-PHY18-20675.
\eject


\begin{thebibliography}{99}

\bibitem{Honerkamp:1996va}
  J.~Honerkamp and K.~Meetz,
  ``Chiral-invariant perturbation theory,''
  Phys.\ Rev.\ D {\bf 3}, 1996 (1971).
  doi:10.1103/PhysRevD.3.1996


\bibitem{Gerstein:1971fm}
  I.~S.~Gerstein, R.~Jackiw, S.~Weinberg and B.~W.~Lee,
  ``Chiral loops,''
  Phys.\ Rev.\ D {\bf 3}, 2486 (1971).
  doi:10.1103/PhysRevD.3.2486

\bibitem{Fradkin:1974df}
E.~S.~Fradkin and G.~A.~Vilkovisky,
``S matrix for gravitational field. ii. local measure, general relations, elements of renormalization theory,''
Phys. Rev. D \textbf{8}, 4241-4285 (1973)
doi:10.1103/PhysRevD.8.4241



\bibitem{Weinberg:1995mt}
S.~Weinberg,
``The Quantum theory of fields. Vol. 1: Foundations,'' (Cambridge University Press, Cambridge, UK, 1995) p. 318.

\bibitem{DeWitt:1962ud}
B.~S.~DeWitt,
``Quantization of fields with infinite-dimensional invariance groups. III. Generalized Schwinger-Feynman theory,''
J. Math. Phys. \textbf{3}, 1073-1093 (1962)
doi:10.1063/1.1703819



\bibitem{Boulware:1970zc}
D.~Boulware,
``Renormalizeability of massive non-abelian gauge fields - a functional integral approach,''
Annals Phys. \textbf{56}, 140-171 (1970)
doi:10.1016/0003-4916(70)90008-4


\bibitem{Salam:1971sp}
A.~Salam and J.~Strathdee,
``Equivalent formulations of massive vector field theories,''
Phys. Rev. D \textbf{2}, 2869-2876 (1970)
doi:10.1103/PhysRevD.2.2869






\bibitem{Charap:1970xj}
J.~Charap,
``Closed-loop calculations using a chiral-invariant lagrangian,''
Phys. Rev. D \textbf{2}, 1554-1561 (1970)
doi:10.1103/PhysRevD.3.1998


\bibitem{Donoghue:1983hi}
  J.~F.~Donoghue, E.~Golowich and B.~R.~Holstein,
  ``Long Distance Chiral Contributions to the $K_L K_S$ Mass Difference,''
  Phys.\ Lett.\  {\bf 135B}, 481 (1984).
  doi:10.1016/0370-2693(84)90320-4



\bibitem{Niedermaier:2006wt}
  M.~Niedermaier and M.~Reuter,
  ``The Asymptotic Safety Scenario in Quantum Gravity,''
  Living Rev.\ Rel.\  {\bf 9}, 5 (2006).
  doi:10.12942/lrr-2006-5

\bibitem{Reuter:1996cp}
  M.~Reuter,
  ``Nonperturbative evolution equation for quantum gravity,''
  Phys.\ Rev.\ D {\bf 57}, 971 (1998)
  doi:10.1103/PhysRevD.57.971
  [hep-th/9605030].


\bibitem{Reuter:2019byg}
  M.~Reuter and F.~Saueressig,
  ``{\it Quantum Gravity and the Functional Renormalization Group : The Road towards Asymptotic Safety},'' (Cambridge University Press, Cambridge, 2018).



\bibitem{Codello:2006in}
  A.~Codello and R.~Percacci,
  ``Fixed points of higher derivative gravity,''
  Phys.\ Rev.\ Lett.\  {\bf 97}, 221301 (2006)
  doi:10.1103/PhysRevLett.97.221301
  [hep-th/0607128].


\bibitem{Donoghue:2019clr}
J.~F.~Donoghue,
``A Critique of the Asymptotic Safety Program,''
Front. in Phys. \textbf{8}, 56 (2020)
doi:10.3389/fphy.2020.00056
[arXiv:1911.02967 [hep-th]].

\bibitem{Codello:2011js}
A.~Codello,
``Large N Quantum Gravity,''
New J. Phys. \textbf{14}, 015009 (2012)
doi:10.1088/1367-2630/14/1/015009
[arXiv:1108.1908 [gr-qc]].





\end{thebibliography}
\end{document}